\documentclass[
  reprint,
  superscriptaddress,
  amsmath,
  amssymb,
  aps,
  pre,
  longbibliography
]{revtex4-2}

\usepackage[utf8]{inputenc}
\usepackage[lining,semibold]{libertine}
\usepackage[libertine,cmintegrals,bigdelims,vvarbb]{newtxmath}

\usepackage{amsmath}
\usepackage{amsfonts}
\usepackage{bm}

\usepackage{graphicx}
\usepackage{array}[=2016-10-06]
\usepackage{dcolumn}
\usepackage{booktabs}
\usepackage{siunitx}

\usepackage{xcolor}
\definecolor{webgreen}{rgb}{0,.5,0}
\definecolor{webbrown}{rgb}{.6,0,0}
\definecolor{RoyalBlue}{rgb}{0.0,0.14,0.4}

\usepackage{hyperref}
\hypersetup{
  colorlinks=true,
  linktocpage=true,
  breaklinks=true,
  pdfstartpage=1,
  pdfstartview=Fit,
  pdfpagemode=UseOutlines,
  plainpages=false,
  bookmarksnumbered=true,
  bookmarksopen=true,
  bookmarksopenlevel=1,
  hypertexnames=true,
  pdfhighlight=/O,
  urlcolor=webbrown,
  linkcolor=RoyalBlue,
  filecolor=RoyalBlue,
  citecolor=webgreen,
  pdftitle={Mutual linearity from classical response theory},
  pdfauthor={Timur Aslyamov and Massimiliano Esposito},
  pdfcreator={pdfLaTeX},
  pdfproducer={LaTeX REVTeX}
}

\usepackage[capitalize]{cleveref}
\AtBeginEnvironment{thebibliography}{\interlinepenalty=10000}

\makeatletter
\def\maketag@@@#1{\hbox{\m@th\normalfont\normalsize#1}}
\makeatother

\begin{document}
\title{Mutual linearity from classical response theory}

\author{Timur Aslyamov}
\email{timur.aslyamov@uni.lu}
\affiliation{Complex Systems and Statistical Mechanics, Department of Physics and Materials Science, University of Luxembourg, 30 Avenue des Hauts-Fourneaux, L-4362 Esch-sur-Alzette, Luxembourg}

\author{Massimiliano Esposito}
\email{massimiliano.esposito@uni.lu}
\affiliation{Complex Systems and Statistical Mechanics, Department of Physics and Materials Science, University of Luxembourg, 30 Avenue des Hauts-Fourneaux, L-4362 Esch-sur-Alzette, Luxembourg}

\date{October 12, 2026}

\begin{abstract}
We unify known mutual-linearity relations for Markov jump processes
and chemical reaction networks within a common framework based on
classical finite-response theory. In a Markov jump process, an exact finite-response
identity shows that all stationary probability changes share a common
factor when the rates of one edge change. Eliminating this
factor gives the relations between probabilities, observables, and
currents, even far from equilibrium. We also explain how the same
argument applies when one coefficient of a linear system changes and
recovers the known relations in Laplace space and between stationary
complex activities in chemical reaction networks.
\end{abstract}
\maketitle

\section{Introduction}
\label{sec:introduction}

Stationary probabilities and currents generally depend nonlinearly on
transition rates. Nevertheless, varying the rates of one edge produces
straight-line relations between these quantities, known as mutual
linearity~\cite{Harunari2024,Khodabandehlou2025,DalCengio2025,Polettini2026Coplanarity,BebonSpeck2026,ZhengLu2026,VoitsSchwarz2026}.
Our aim is to clarify the common origin of these relations and provide
a concise derivation using the Drazin inverse. We start from the
classical finite-response identity~\cite{Schweitzer1968,Meyer1980}.
When the rates of one edge change, this identity shows that all
stationary probability changes share a common factor.
Eliminating it gives the relations between probabilities,
observables, and currents. This approach gives short derivations of
the stationary and Laplace-space
relations~\cite{Harunari2024,ZhengLu2026,VoitsSchwarz2026} and the
relations between stationary complex activities in chemical reaction
networks~\cite{HarunariFiusaPolettini2026}. We also explain when the
same property holds in other linear systems.

We focus on the common derivation of these relations. The works cited above also establish more detailed results beyond the scope of this paper. 
These include explicit network formulas for the coefficients relating currents~\cite{Harunari2024}, expressions for these coefficients
in terms of mean first-passage times~\cite{Khodabandehlou2025}, relations
for several input currents~\cite{DalCengio2025}, bounds on stationary
probabilities and observables~\cite{BebonSpeck2026}, and relations between
Laplace transforms of first-hitting-time distributions~\cite{VoitsSchwarz2026}.

\section{Exact stationary response}
\label{sec:response}

The finite-response identity shown below dates back to the
1960s~\cite{Schweitzer1968} and was originally formulated for Markov chains. Here we write it for continuous-time Markov jump processes on a finite state space.

We consider a finite irreducible Markov jump process, $\partial_t\bm p=W\bm p$, where $W_{ij}$ is the transition rate from $j$ to $i$, and $W_{jj}=-\sum_{i\ne j}W_{ij}$.
The stationary distribution satisfies $W\bm\pi=0$ and $\bm1^\intercal\bm\pi=1$.
For a second irreducible generator $W'=W+\Delta W$, the classical fundamental-matrix identity takes the form
\begin{align}
    \Delta\bm\pi\equiv\bm\pi'-\bm\pi=-G\Delta W\bm\pi'\,,
    \label{eq:finite}
\end{align}
where $G=W^D$ is the Drazin inverse~\cite{Meyer1975,crook2018drazin,aslyamov2025excess}.
It satisfies $GW=WG=I-\bm\pi\bm1^\intercal$, $G\bm\pi=0$, and $\bm1^\intercal G=0$.
Indeed, subtracting the two stationary equations gives $W\Delta\bm\pi=-\Delta W\bm\pi'$.
Multiplication by $G$ and normalization give \cref{eq:finite}, without assuming that $\Delta W$ is small. 

The Drazin inverse underlies current-cumulant
calculations~\cite{BaiesiMaesNetocny2009}, expansions under slow
driving~\cite{MandalJarzynski2016}, fluctuation-response relations for
state observables~\cite{Ptaszynski2026State} and state--current
correlations~\cite{Ptaszynski2026Mixed}, responses in the Curie--Weiss
model~\cite{PtaszynskiEsposito2025}, and finite perturbations of Markov jump
processes~\cite{BaoLiang2026}. The relation
$\tau_{j\to i}=(G_{ij}-G_{ii})/\pi_i$
connects this approach to mean first-passage
times~\cite{ChoMeyer2000,Harvey2023,Khodabandehlou2025,BaoLiang2026}.
Alternatively, normalization gives an invertible matrix from
the rate matrix~\cite{AslyamovEsposito2024Exact,AslyamovEsposito2024General}.

\section{Mutual linearity}
\label{sec:mutual-linearity}

\subsection{State observables}

For the second matrix $W'$, we change $W_{ij}$ and $W_{ji}$,
keeping all other non-diagonal rates fixed and adjusting the diagonal elements to preserve probability conservation.
Then, only components $i$ and $j$ of the vector $\Delta W\bm\pi'$ can be nonzero, and they sum to zero. Equation~\eqref{eq:finite} therefore gives
\begin{align}
    \Delta\pi_n
    &=(G_{nj}-G_{ni})
      (\Delta W_{ij}\pi_j'-\Delta W_{ji}\pi_i')
    \nonumber\\
    &\equiv h_n\alpha\,,
    \label{eq:edge-response}
\end{align}
where $h_n\equiv G_{nj}-G_{ni}$ is calculated from the rates before the change
and does not depend on $\Delta W_{ij}$ or $\Delta W_{ji}$.
All dependence on these rate changes enters the second factor
$\alpha \equiv \Delta W_{ij}\pi_j'-\Delta W_{ji}\pi_i'$.
Repeating \cref{eq:edge-response} for state $m$ and eliminating $\alpha$ gives
\begin{align}
    \Delta\pi_n=\frac{h_n}{h_m}\Delta\pi_m\,,
    \qquad h_m\ne0\,.
    \label{eq:probabilities}
\end{align}
Thus, $(\pi_m',\pi_n')$ lies on one straight line for every choice of
the new edge rates. This property was recently discovered in Ref.~\cite{BebonSpeck2026}.

More generally, for state observables $O=\sum_n O_n\pi_n$ and
$Q=\sum_n Q_n\pi_n$ with fixed weights, the same argument gives
\begin{align}
    \Delta O=\frac{\sum_n O_n h_n}{\sum_n Q_n h_n}\Delta Q\,,
    \label{eq:observables}
\end{align}
provided that the denominator is nonzero.

Let us now consider the linear-response limit of \cref{eq:edge-response}:
\begin{align}
    \partial_{W_{ij}}\pi_n=h_n\pi_j\,,\qquad
    \partial_{W_{ji}}\pi_n=-h_n\pi_i\,,
    \label{eq:derivatives}
\end{align}
so that $-\partial_{W_{ij}}\pi_n/\partial_{W_{ji}}\pi_n=\pi_j/\pi_i$
whenever $h_n\ne0$. This recovers the response ratio of
Ref.~\cite{BebonSpeck2026}.
Interestingly, our dynamical fluctuation-response theory~\cite{AslyamovEsposito2026} gives the same ratio for
dynamical responses to impulsive rate perturbations at time $t$, with
$\pi_j/\pi_i$ replaced by $p_j(t)/p_i(t)$.
Thus, this response ratio can be obtained without invoking mutual linearity under finite rate changes.

\subsection{Currents}

The stationary current from $l$ to $k$ is
\mbox{$J_{kl}\equiv W_{kl}\pi_l-W_{lk}\pi_k$}.
For an edge whose transition rates $W_{kl}$ and $W_{lk}$ are kept fixed,
\cref{eq:edge-response} gives
\begin{align}
    \Delta J_{kl}=(W_{kl}h_l-W_{lk}h_k)\alpha\,,
    \label{eq:output-current}
\end{align}
where only the change in the stationary distribution contributes.
On the perturbed edge, the rates themselves also change:
\begin{align}
    \Delta J_{ij}
    &=W_{ij}\Delta\pi_j-W_{ji}\Delta\pi_i
      +\Delta W_{ij}\pi_j'-\Delta W_{ji}\pi_i'
    \nonumber\\
    &=(1+W_{ij}h_j-W_{ji}h_i)\alpha\,.
    \label{eq:input-current}
\end{align}
Eliminating $\alpha$ between the currents on $(i,j)$ and another edge
$(k,l)$ gives
\begin{align}
    \Delta J_{kl}
    =\frac{W_{kl}h_l-W_{lk}h_k}
          {1+W_{ij}h_j-W_{ji}h_i}\Delta J_{ij}\,,
    \label{eq:currents}
\end{align}
which is the mutual linearity of stationary
currents~\cite{Harunari2024,Khodabandehlou2025,DalCengio2025,Polettini2026Coplanarity}.
Both current changes contain the same $\alpha$, so the coefficient relating them is independent of the new edge rates, provided the denominator in \cref{eq:currents} is nonzero.


\section{Further applications}
\label{sec:applications}

We now apply the common-factor argument to linear systems, Laplace-transformed dynamics
and stationary complex activities. 

\subsection{Linear systems} 
First, consider a linear
system $A\bm x=\bm b$ with nonsingular $A$ and fixed $\bm b$.
If only $A_{ij}$ changes and $A'$ is nonsingular, subtracting the
equations and eliminating $\Delta A_{ij}x_j'$ between $n$ and $m$ gives
\begin{align}
    \Delta x_n&=-(A^{-1})_{ni}\Delta A_{ij}x_j'\,,\nonumber\\
    \Delta x_n&=\frac{(A^{-1})_{ni}}{(A^{-1})_{mi}}\Delta x_m\,,
    \qquad (A^{-1})_{mi}\ne0\,.
    \label{eq:linear-system}
\end{align}
The same proportionality holds for fixed linear observables. Moreover, this argument applies whenever $\Delta A\bm x'=\bm u\alpha$,
with fixed $\bm u$: the perturbation enters the equations in fixed
proportions, with only its scalar amplitude $\alpha$ changing.
Then $\Delta\bm x=-A^{-1}\bm u\alpha$, so eliminating $\alpha$
again gives mutual linearity.

\subsection{Laplace-transformed dynamics}

The frequency-domain extension follows from this argument using $A=sI-W$. For time-independent rates, the Laplace-transformed master equation is
$(sI-W)\widetilde{\bm p}(s)=\bm p(0)$.
We keep the same initial distribution in both systems and define the resolvent matrix
$R(s)=(sI-W)^{-1}$ for $\operatorname{Re}s>0$.
Changing the rates of edge $(i,j)$ and eliminating the common factor
between states $n$ and $m$ results in
\begin{align}
    \Delta\widetilde p_n
    &=h_n(s)(\Delta W_{ij}\widetilde p_j'
              -\Delta W_{ji}\widetilde p_i')\,,\nonumber\\
    \Delta\widetilde p_n
    &=\frac{h_n(s)}{h_m(s)}\Delta\widetilde p_m\,,
    \qquad h_m(s)\ne0\,,
    \label{eq:laplace}
\end{align}
where $h_n(s)=R_{ni}(s)-R_{nj}(s)$ and all transformed probabilities are evaluated at the same $s$.
This recovers the Laplace-domain probability relations~\cite{ZhengLu2026,VoitsSchwarz2026}.
Repeating \cref{eq:output-current,eq:input-current,eq:currents} with
$\widetilde{\bm p}(s)$ and $h_n(s)$ gives the corresponding current
relations~\cite{Harunari2024}.
These relations hold for the Laplace transforms. At a fixed time, varying the edge rates need not produce a straight-line relation between two probabilities or two currents~\cite{VoitsSchwarz2026}.

\subsection{Chemical reaction networks}

The same argument recovers the mutual linearity of stationary complex
activities derived in Ref.~\cite{HarunariFiusaPolettini2026}.
A \emph{complex} is a combination of species appearing on either side
of a reaction. For example, $A+B\rightleftharpoons C$ connects the
complexes $A+B$ and $C$. If $x_A,x_B,x_C$ are the species
concentrations, the corresponding \emph{complex activities} are
$\psi_{A+B}=x_Ax_B$ and $\psi_C=x_C$.

We represent the complexes as graph vertices and the reactions as
edges, and consider a connected mass-action network with bidirectional
reactions.
The matrix $K$ plays the same role as $W$ above: for $i\ne j$,
$K_{ij}$ is the rate constant for the reaction $j\to i$, and each
column sums to zero.
We assume that the positive stationary states considered below are complex
balanced~\cite{MarehalliSrinivas2023}: at each complex, the total
incoming reaction flux equals the total outgoing flux, so
$K\bm\psi=0$.
This equation is linear in the activities, even though the activities
can depend nonlinearly on the concentrations. Unlike probabilities,
the activities are not normalized: their sum $S=\sum_n\psi_n$ can
change with the rate constants. We define the normalized activity
vector $\bm\pi=\bm\psi/S$.

Now change only the forward and backward rate constants $r_+$ and
$r_-$ of one reaction $j\rightleftharpoons i$. As in the derivation
of \cref{eq:finite}, subtracting the two complex-balance equations
gives 
\begin{align}
    K\Delta\bm\psi=-\Delta K\bm\psi'\,.
    \label{eq:complexes-pert}
\end{align}
Multiplying \cref{eq:complexes-pert} by $G=K^D$, with
$GK=I-\bm\pi\bm1^\intercal$, and using
$\bm1^\intercal\Delta\bm\psi=\Delta S$ gives
\begin{align}
    \Delta\bm\psi-\bm\pi\Delta S=-G\Delta K\bm\psi'\,.
\end{align}
For the controlled reaction, this yields
\begin{align}
    \Delta\psi_n&=\pi_n\Delta S+h_n\alpha\,,\nonumber\\
    \alpha&\equiv\Delta r_+\psi_j'-\Delta r_-\psi_i'\,,
    \label{eq:complex-response}
\end{align}
where $h_n=G_{nj}-G_{ni}$ and $\Delta S=S'-S$.
The first term accounts for the change in the total activity; it
vanishes for probabilities because normalization fixes $S=1$.
Equivalently, $\bm\psi'=S'\bm\pi+\bm h\alpha$.
Eliminating the two scalars $S'$ and $\alpha$ between three complexes
$a,b,c$ gives
\begin{align}
    C_{bc}\psi_a'+C_{ca}\psi_b'+C_{ab}\psi_c'=0\,,
    \label{eq:complex-activities}
\end{align}
where $C_{ab}=\pi_a h_b-\pi_b h_a$ is evaluated before the change
and is independent of $\Delta r_+$ and $\Delta r_-$.
Equation~\eqref{eq:complex-activities} recovers the homogeneous
relation in Eq.~(5) of Ref.~\cite{HarunariFiusaPolettini2026}.

\section{Conclusion}
\label{sec:conclusion}

The classical finite-response identity~\eqref{eq:finite} unifies
mutual linearity for stationary probabilities, Eq.~\eqref{eq:probabilities};
state observables, Eq.~\eqref{eq:observables}; currents, Eq.~\eqref{eq:currents};
components of linear systems, Eq.~\eqref{eq:linear-system};
Laplace-transformed probabilities, Eq.~\eqref{eq:laplace}; and stationary
complex activities in complex-balanced chemical reaction networks,
Eq.~\eqref{eq:complex-activities}.

\bibliography{references}
\end{document}